\documentclass[prb,showpacs,twocolumn,amssymb,amsmath]{revtex4}
\usepackage{graphicx,bm} 
\begin{document}
\title{Anisotropic magnetoresistance contribution to measured domain wall 
resistances of in-plane magnetised (Ga,Mn)As}
\author{H.G. Roberts}
\author{S. Crampin}
\author{S. J. Bending}
\affiliation{Department of Physics, University of Bath, Bath BA2 7AY, UK}
\date{\today}

\begin{abstract}
We demonstrate the presence of an important anisotropic magnetoresistance
contribution to the domain wall resistance recently measured
in thin-film (Ga,Mn)As with in-plane magnetic anisotropy. Analytic results
for simple domain wall orientations supplemented by numerical results for
more general cases show this previously omitted contribution can largely
explain the observed negative resistance.
\end{abstract}

\maketitle


The electrical resistance associated with current flow across a domain wall
(DW) separating uniformly magnetised regions in a magnetic material has been 
the subject of investigation since the 1930s. A large volume of
often contradictory results exists, mostly attributable to the difficulty 
in separating normally small intrinsic effects from the myriad of extrinsic 
effects that 
also contribute in magnetoresistance measurements.\cite{Marrows} 
In recent years some consensus has emerged in
the study of DWs in metallic epitaxial films and 
nanostructures\cite{metals} where it is possible to
more fully characterise and control the magnetic microstructure,
with many results consistent with 
the spin-mistracking models of DW resistance.\cite{Viret_PRB,LZ}
Although the small magnitude of the DW resistance limits potential applications,
significant enhancements at 
nanoconstrictions\cite{GarciaBruno} allied with advances 
in the atomic-scale control of materials,
raise hopes of practical magnetoresistive devices,
whilst a close relationship with the phenomenon of
current-induced DW motion\cite{Tatara,Yamanouchi} also makes the understanding 
of DW resistance of considerable importance.

It has been noted that enhanced magnetoresistance effects may be associated 
with DWs in ferromagnetic semiconductors due to the longer Fermi
wavelength\cite{Tatara} and
large exchange splitting relative to band width.\cite{Vignale} 
R\"uster {\it et al.}\cite{Ruester} have observed
an 8\% increase in the magnetoresistance due to DWs pinned at $< 10$\ nm
constrictions in in-plane magnetised (Ga,Mn)As nanostructures, and 
Chiba {\it et al.}\cite{Chiba} report
a significant positive DW magnetoresistance in perpendicularly 
magnetised (Ga,Mn)As layers consistent with the theory 
of Levy and Zhang.\cite{LZ} Tang {\it et al.} \cite{Tang_Nature} 
have studied the resistance of $30-100\ \mu$m devices patterned 
from in-plane magnetised (Ga,Mn)As epilayers. By 
measuring the average resistance $\langle R \rangle$ along the sides of the
device channel as a $90^\circ$ domain wall is driven through by current pulses, 
they find a small resistance \emph{drop}.
Scaling by a wall width of 10\ nm, Tang {\it et al.} deduce a DW 
resistivity as
large as $\Delta\varrho/\rho\sim -100$\% --- a remarkable result implying
resistance free current transport through the region occupied by
the DW. Although theories exist that predict a negative intrinsic DW resistance,
either as a result of modifications to quantum interference 
phenomena\cite{NR_theory} or differences in spin-dependent 
relaxation rates,\cite{Gorkom} this result is many orders of magnitude 
greater than any negative DW resistivity previously reported in a metal.\cite{Marrows}

In this work we demonstrate the existence of a sizeable \emph{extrinsic} 
contribution to the negative DW resistance measured using the experimental 
configuration employed in Ref.  [\onlinecite{Tang_Nature}]. This 
anisotropic magnetoresistance (AMR) effect arises from the circulating
currents induced by the abrupt change in the off-diagonal resistivity 
at $90^\circ$ DW, and persists even after the resistance is 
averaged across the sample. An analytic expression is derived for the
simplest DW orientation, supplemented by numerical results for the more
general case, which also allow us to simulate the experiments where a 
current-driven DW is moved through a microdevice.

\begin{figure}[b]
\includegraphics[width=65mm]{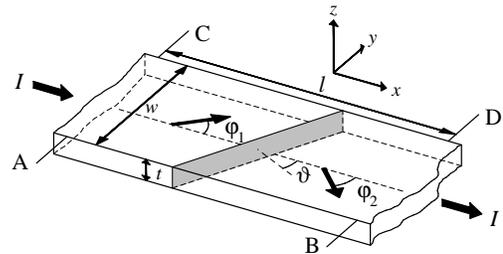}
\caption{The geometry considered in this work. A domain wall 
inclined at an angle $\vartheta$ 
separates regions in which the in-plane magnetisation lies at angles 
$\varphi_1$ and $\varphi_2$. Potential measurements are made using
probes A, B, C and D.}
\label{fig:geometry}
\end{figure}

We consider the current flow within an infinitely long thin conducting sample
with rectangular cross section, width $w$, thickness $t$. The sample lies
parallel to the $xy$ plane with the long edge parallel to the $x$ axis 
(Fig. \ref{fig:geometry}). A dc electrical current $I$ flows through the 
sample. Ideal probes are attached and measure the potential at points 
on either side of the sample separated by a distance $l$.

The current density $\bm{J}(x,y)$ is assumed to be uniform as 
$x\rightarrow\pm\infty$, and no current flows through the sides of the device:
\begin{subequations}
\label{eqn:j_bcs}
\begin{eqnarray}
\bm{J}(\pm\infty,y)&=&(I/wt,0)=(j,0)\label{eqn:j_bcsa}\\
J_y(x,0)&=&J_y(x,w)=0.\label{eqn:j_bcsb}
\end{eqnarray}
\end{subequations}
Within the sample, $\bm{J}$ is found by satisfying current continuity 
and the steady state Maxwell equation,
with the electric field $\mathbf{E}$ and current density related via Ohm's law
\begin{subequations}
\label{eqn:je_eqns}
\begin{eqnarray}
\nabla\cdot\mathbf{J}&=&0 \label{eqn:je_eqns_a} \\
\nabla\times\mathbf{E}&=&0 \label{eqn:je_eqns_b}\\
\bm{E}&=&\hat{\bm{\rho}}\bm{J} 
\label{eqn:ohm}
\end{eqnarray}
\end{subequations}
with $\hat{\bm{\rho}}$ a spatially varying resistivity tensor.
With in-plane magnetised (Ga,Mn)As epilayers, the resistivity
in directions parallel ($\rho_\parallel$) 
and perpendicular ($\rho_\perp$) to the magnetisation 
differ\cite{Tang_PRL}
with $\rho_\parallel < \rho_\perp$.
If $\varphi$ is the magnetisation direction in a given domain 
(Fig. \ref{fig:geometry}), the corresponding cartesian tensor is
\begin{eqnarray}
{\hat{\bm{\rho}}} &=& R_\varphi^{-1}\left(
\begin{array}{cc}
\rho_\parallel & 0 \\
0              & \rho_\perp
\end{array}\right) R_\varphi\nonumber
\\  &=& \bar{\rho} \left(
\begin{array}{cc}
1+\frac{\beta}{2}\cos 2\varphi & \frac{\beta}{2}\sin 2\varphi \\
\frac{\beta}{2}\sin 2\varphi   & 1-\frac{\beta}{2}\cos 2\varphi
\end{array}
\right)
\label{eqn:rho_gamnas}
\end{eqnarray}
where $\bar{\rho}=(\rho_\perp+\rho_\parallel)/2$, 
$\beta=(\rho_\parallel-\rho_\perp)/\bar{\rho}$.
With cubic anisotropy, domain walls divide regions which have magnetisation
directions differing by $90^\circ$, but in the experiments recently reported 
this is modified by a weak in-plane uniaxial anisotropy. We denote by 
$\vartheta$ the angle which the normal to the wall $(\hat{\bm{n}}_{\rm DW})$
makes with respect to the $x$ axis.
Fig. \ref{fig:geometry} shows the geometry in the case of a single 
domain wall within the device. 

At the wall itself, the boundary conditions are continuity in the normal
component of the current and in the tangential component of the 
electric field:
\begin{subequations}
\label{eqn:bcs}
\begin{eqnarray}
\label{eqn:bc_J}
{\bf J}\perp\hat{\bm{n}}_{\rm DW}&=&J_x\cos\vartheta+J_y\sin\vartheta\\
{\bf E}\parallel\hat{\bm{n}}_{\rm DW}&=&-E_x\sin\vartheta+E_y\cos\vartheta\nonumber\\
&=&-(\varrho_{xx}J_x+\varrho_{xy}J_y)\sin\vartheta\nonumber\\
&&+(\varrho_{yx}J_x+\varrho_{yy}J_y)\cos\vartheta.
\label{eqn:bc_E}
\end{eqnarray}
\end{subequations}
Matching (\ref{eqn:bc_E}) when the resisitivity tensor elements are different 
on either side of the DW is not possible with a uniform current 
$\bm{J}=(j,0)$; the DW induces circulating currents and it is these 
that give rise to an AMR contribution to the resistance across the wall.

To see this, we first consider the case where the device channel contains a 
DW in the $yz$ plane ($\vartheta=0$) at $x=x_0$, separating regions in which
the in-plane magnetisation is at an angle $\varphi_1$ in region 1 
and $\varphi_2$ in 
region 2. The average of the longitudinal resistances measured along
opposite sides of the device, used in Ref. [\onlinecite{Tang_Nature}]
to eliminate contributions from the planar Hall effect, 
can be expressed in terms of a 
difference in voltages at either end of the device channel:
\begin{eqnarray}
-I\langle R \rangle&=&(V_B-V_A)/2+(V_D-V_C)/2\nonumber\\
&=&(V_B+V_D)/2-(V_A+V_C)/2.
\label{eqn:vdiff}
\end{eqnarray}
(The minus sign is because the potential \textit{falls} along the direction
of positive current flow.)
We take the length $l$ (and $x_0$) to be large enough so that the
static eddy currents induced by the DW are fully contained within the area 
of the device defined by the 4 probes, and then the current has its 
asymptotic value at both $x=0$ and $x=l$. This means that the electric field in 
the $y$-direction
is constant both between C and A, 
$E_y(0,y)=\rho_{yx}^1 j$, and D and B, $E_y(l,y)=\rho_{yx}^2 j$,
and the voltage changes linearly
between points C and A, and between D and B.
The voltage averages in
(\ref{eqn:vdiff}) can then be re-expressed as integrals of the voltage
\emph{across} the device, e.g.
\begin{equation}
(V_B+V_D)/2=\frac{1}{w}\int_{0}^{w} V(l,y) dy,
\end{equation}
and the two terms in
(\ref{eqn:vdiff}) then can be combined using
$V(l,y)-V(0,y)=-\int_0^l E_x(x,y) dx$
to give $\langle R \rangle$ in terms of an integral of the electric field
over the area of the device between the probes:
\begin{equation}
I\langle R \rangle = \frac{1}{w}\int_{x=0}^l\int_{y=0}^w\! E_x(x,y)dxdy.
\end{equation}
Splitting the integral into separate contributions from the two domains
within each of which the resistivity tensor is constant, and using the
following results
\begin{equation}
\int_0^wJ_x(x,y)dy=jw,\quad
\int_0^lJ_y(x,y)dx=0
\end{equation}
that are found by integrating the continuity equation over regions
$\Omega=\{(x',y)\in \mathbb{R}^2\ |\ 0\le x' \le x, 0 \le y \le w\}$
and 
$\Omega=\{(x,y')\in \mathbb{R}^2\ |\ 0\le x \le l, 0 \le y' \le y\}$
respectively, yields
\begin{eqnarray}
\!\!\!\!\!\!\!\!\langle R \rangle&\!=\!&
\frac{1}{wt}\left[\rho_{xx}^1 x_0 +\rho_{xx}^2 (l-x_0)\right]\nonumber\\
&&+\frac{1}{jw^2t}\left(\rho_{xy}^1-\rho_{xy}^2\right)
\int_{x=0}^{x_0}\int_{y=0}^w\!J_y(x,y)dxdy.
\label{eqn:res_final}
\end{eqnarray}
The first term describes a resistance that linearly interpolates 
between the asymptotic resistances of the channel in the two
uniform magnetisation states. In Ref. [\onlinecite{Tang_Nature}] differences
between measured resistance values and this linear interpolation have been 
interpreted as originating from an intrinsic DW resistivity. 
However, the final term in (\ref{eqn:res_final}), which henceforth we denote
$R_{\mathrm{AMR}}$, is a new
contribution that we find, which results directly from the discontinuity in 
the resisitivity at the DW
and which is proportional to the total parallel current induced
on either side of the DW.

To obtain an estimate for the value
of this additional contribution to the DW resistance, we consider 
the case where $\varphi_1=-\varphi_2=\varphi$ which applies when the hard 
axis is
perfectly aligned along the device channel. Then the diagonal components of the
resistivity tensors are continuous across the DW, and the off-diagonal
components change sign. For small $\beta$ the longitudinal current component
$J_x$ will be dominated by the uniform background current $j$, except within a 
distance $\sim\beta w$ of the sides of the device near the DW where the 
current perturbation is concentrated. Neglecting this edge correction,
it then follows from Eqn. (\ref{eqn:bcs}) that
immediately on either side of the DW
\begin{equation}
J_y(x_0\pm 0^+,y)\simeq\pm \frac{j\beta}{2}\sin 2\varphi.
\label{eqn:jy_approx}
\end{equation}
Ignoring anisotropy, the slowest decaying current perturbations decay 
like\cite{bate} $\exp -\pi|x|/w$ and 
by assuming that $J_y$ decays like this from the interface value 
(\ref{eqn:jy_approx}) we can evaluate the integral in (\ref{eqn:res_final}) 
to get for the AMR contribution to
the average longitudinal resistance
\begin{equation}
R_{\mathrm{AMR}}=-R_\square
\frac{\beta^2}{2\pi}\sin^22\varphi
\label{eqn:r_amr}
\end{equation}
where $R_\square =\bar{\rho}/t$ is the sheet resistance.
This result shows how the circulating currents give rise to a negative
contribution to the resistance across the DW.

\begin{figure}[t]
\includegraphics[width=55mm]{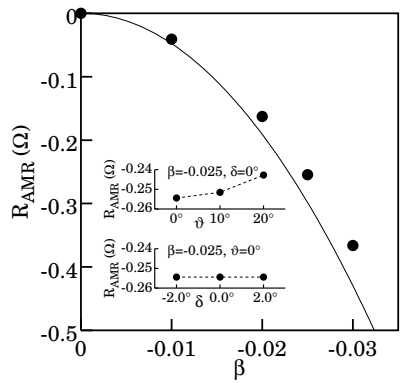}
\caption{Calculated AMR contribution to the resistance across a DW
for different values of the anisotropy $\beta$. Symbols, 
numerically calculated values; solid line, values using Eqn. (\ref{eqn:r_amr}). 
A film thickness $t=100$\ nm,
resistivity $\bar\rho=3\times 10^{-4}\ \Omega\textrm{m}$,
misalignment $\delta=0^\circ$, DW angle $\vartheta=0^\circ$ and
magnetisation angle $\varphi=45^\circ$ have been used.
Insets: the effect of varying the DW angle $\vartheta$ and 
misalignment $\delta$.}
\label{fig:R_vs_beta}
\end{figure}

We have also performed numerical studies of the current distributions and 
resulting fields and voltages in the presence of DWs. 
The numerical solution is
not restricted to the specific configuration that was assumed in deriving
the analytic estimate for $R_{\mathrm{AMR}}$, so as well
as enabling an assessment of the accuracy of this expression obtained
we are also able to include the effects of misalignment of the hard axis 
with respect to the device channel, and the angle of the domain wall.
Some results are given in Fig. \ref{fig:R_vs_beta}.
The solution is obtained by introducing a stream function $\psi(x,y)$ 
that is related to the current density via 
${\bf J}=(\partial \psi/\partial y,-\partial \psi/\partial x)$, thereby
ensuring
that current continuity Eqn. (\ref{eqn:je_eqns_a}) is 
satisfied. Combining Eqns (\ref{eqn:je_eqns_b}) and (\ref{eqn:ohm}) then
results in a non-separable elliptic partial differential equation\cite{tang}
that we solve for $\psi$ via the multigrid relaxation method. 

\begin{figure}[t]
\includegraphics[width=65mm]{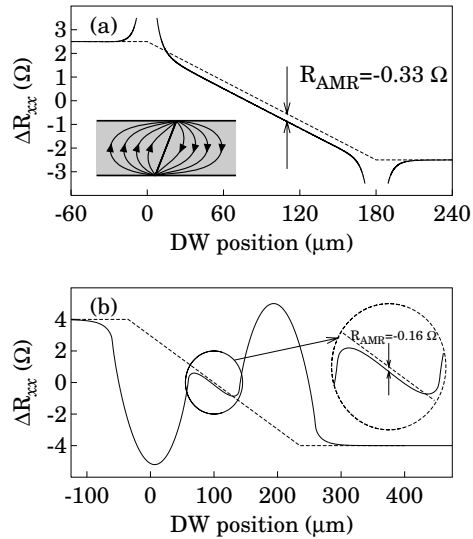}
\caption{(a) Variation in average resistance along the sides of the device
channel (between $x=0$ and $x=180\ \mu$m) as a function of the position of 
the DW. The aspect ratio is $l/w=6$ and the DW inclined at $\vartheta=20^\circ$.
See the text for the other parameters used. The inset shows the induced 
current flow.
(b) Similar to (a) but for different material parameters (see text)
and dimensions: the resistance is measured between $x=0$ and 
$x=200\ \mu$m, $l/w=2$ and $\vartheta=50^\circ$.
}
\label{fig:R_vs_x}
\end{figure}

In Fig. \ref{fig:R_vs_x}a we show typical results from our numerical studies,
displaying the variation in the longitudinal resistance
as a DW inclined at an angle $\vartheta=20^\circ$ passes along a device 
channel of width $w=30\mu$m and with voltage 
probes separated by $l=180\mu$m. The average resistance of the two 
uniform magnetisation states has been subtracted:
$\Delta R_{xx}=\langle R \rangle -(\varrho_{xx}^1+\varrho_{xx}^2)l/(2wt)$.
A general linear variation in the resistance is seen,
except when the probes are within the range of the 
circulation currents induced by the DW; these cause a rapid
variation over a distance $\sim w\tan\vartheta+2w/\pi$
as expected from geometrical considerations and the discussion 
following Eq. (\ref{eqn:jy_approx}). 
Furthermore, we see that the calculated resistance lies \emph{below} 
a straight line interpolation performed between the two asymptotic channel
resistances of the two magnetisation states. In this calculation
we use values that correspond as best as possible 
to the system reported in Fig. 4 of Ref. 
[\onlinecite{Tang_Nature}]:
film resistivity $\bar{\varrho}=3\times 10^{-4} \ \Omega$m, 
thickness $t=100$\ nm, and anisotropy\cite{beta} $\beta=-0.03$. 
The magnetisation
orientations within the two domains are taken to be 
$\phi_1=\delta+\phi,\phi_2=\delta-\phi$ where $\phi=37^\circ$ due to uniaxial 
anisotropy\cite{Tang_PRL,Tang_PRB} and the misalignment
$\delta=-0.28^\circ$
(the difference between the asymptotic resistances,
\begin{equation}
(\varrho_{xx}^2-\varrho_{xx}^1)\frac{l}{wt}=
\frac{\bar{\varrho}\beta l}{wt}\sin 2\phi\sin 2\delta,
\end{equation}
is then $5\ \Omega$ as found in Ref. [\onlinecite{Tang_Nature}]). 
Using these values,
numerically we find the resistance is lowered by $0.33\ \Omega$ as a result of
the eddy currents induced by the DW.

In Fig. \ref{fig:R_vs_beta} we compare numerical values for 
$R_{\mathrm{AMR}}$ found in a number of similar calculations to 
that just described, with those obtained using Eqn. (\ref{eqn:r_amr}).
The numerical results also display the $\beta^2$ dependence and,
as expected given the approximations made in estimating the current
integral, our analytic expression  overestimates the actual resistance,
we find by approximately 15\% when $\vartheta=0$. 
This value of $R_{\mathrm{AMR}}$ is further reduced at DWs 
inclined relative to the 
current direction, but is relatively insensitive to the misalignment angle
(insets in Fig. \ref{fig:R_vs_beta}). Thus Eqn. (\ref{eqn:r_amr}) has
some value in estimating the AMR contribution to the DW resistance,
but numerical calculations are required for accurate estimates.

In Fig. \ref{fig:R_vs_x}b we show the calculated longitudinal resistance for
a second case, with parameters chosen to correspond to the device 
reported in Fig. 1 in Ref. [\onlinecite{Tang_Nature}]. This was the initial 
device studied experimentally, in which the misalignment of the hard axis 
with the device channel is greater.
We use $\bar{\varrho}=4\times 10^{-4}\ \Omega$m, $t=150$\ nm, $\beta=-0.03$,
with the channel width $w=100\ \mu$m and voltage probes separated by 
$200\ \mu$m. Also, $\phi=-37^\circ$, $\delta=1.5^\circ$.
The greater structure exhibited by $\Delta R_{xx}$ in this case is due to 
a larger DW inclination ($\vartheta=50^\circ$) and the smaller aspect ratio 
$l/w$ of the device, and results in a linear variation over only a short
range of DW positions midway between the voltage probes. The precise results 
are rather sensitive to the value of $\vartheta$. However, generally we find 
that in the linear region the large spatial extent of the eddy currents still 
affects the slope of the resistance curve, which no longer
coincides with a linear interpolation of the asymptotic resistances between
the probe positions. The dashed line in Fig. \ref{fig:R_vs_x}b which is 
parallel to the linear section of the resistance curve connects points some 
30\% further apart than the probes. The calculated resistance is again
lowered due to the AMR, but the difference of $-0.16\ \Omega$ is smaller 
than the value ($-0.18\ \Omega$) found if the distance between the
voltage probes is increased so as to fully contain the eddy currents.

The magnetisation profile \textit{within} the wall can also give rise to
a negative AMR. However, the contribution we describe above dominates here.
Assuming a $90^\circ$ N\'eel like 
wall with magnetisation rotating like $\phi(x)=-(1/2)\tan^{-1}\sinh x/\lambda$,
where $\pi\lambda$ is the wall width,
gives a contribution to leading order of $\lambda/(-2\beta w) R_{\rm AMR}$,
or just a few percent of the contribution from the circulation currents.
Other wall profiles in which the spin rotates out of the plane 
lead to the same conclusion. Only if the DWs in this system were $180^\circ$
walls would the in-wall contribution be significant, since then the circulation
current contribution (\ref{eqn:r_amr}) vanishes.

Comparing with experiment, the DW resistances reported in 
[\onlinecite{Tang_Nature}] for the devices 
modelled in Figs \ref{fig:R_vs_x}a and \ref{fig:R_vs_x}b are 
$-1.0\pm 0.2\ \Omega$ and $-0.44\pm 0.5\ \Omega$ respectively; a third set of 
devices similar to that of Fig. \ref{fig:R_vs_x}a but with $w=60\ \mu$m
gave $-0.3\pm 0.2\ \Omega$. The corresponding $R_{\rm AMR}$ values we find are
$-0.33\ \Omega$, $-0.16\ \Omega$ and $-0.33\ \Omega$. The 
previously neglected AMR contributions to the resistance across the DW
make a major contribution to, and can largely explain, the negative values 
observed, with the exception of one set of devices where a true negative 
intrinsic DW resistance may indeed have been observed.
Clearly further experiments are required to clarify the situation,
before attempts to quantitatively account for the DW resistance\cite{osz}
can be properly assessed. For these,
devices with a large aspect ratio $l/w$, and containing DWs orientated
normal to the device channel, are clearly desirable. 

To summarise, we have identified a significant anisotropic magnetoresistance 
contribution to the negative domain wall resistivities recently observed
in microdevices fabricated from (Ga,Mn)As epilayers. We derive an analytic 
estimate of the magtitude of this contribution, and report calculations of the 
channel resistance as a DW is moved through the device which provide a
good description of the experiments.

This work was supported by the Leverhulme Trust, through Grant No. F/00\ 351 F.


\begin{thebibliography}{}
\bibitem{tang}A similar approach was used in [\onlinecite{Tang_PRB}] but the 
series solution then obtained does not satisfy the original equation.
\bibitem{beta} We determine $\beta$ from the quoted values of the planar Hall 
resistance $R_{xy}$, using $\beta=-2R_{xy}t/\bar{\varrho}\sin2\phi$.
\end{thebibliography}
\end{document}